\UseRawInputEncoding
\documentclass[journal=jacsat,manuscript=article]{achemso}

\usepackage{subcaption}
\usepackage{xcolor}
\usepackage[version=3]{mhchem} 

\author{Xiaochen Jin}
\author{Shunda Chen}
\author{Tianshu Li}
\email{tsli@email.gwu.edu}
\affiliation[George Washington University]
{Department of Civil and Environmental Engineering, George Washington University,
Washington, DC 20052}

\title
  {Coexistence of two types of short-range order in SiGeSn medium-entropy alloys}


\begin{document}

\begin{abstract}

Short-range chemical order (SRO) has been recently demonstrated to play a decisive role in modulating a wide range of physical properties in medium-entropy alloy (MEA) and high-entropy alloy (HEA). The enormous configurational space of these alloys implies multiple forms of SRO are likely to develop concurrently but such structural diversity has not been reported. Here we show, through extensive {\em ab initio}-based sampling study, that SiGeSn medium-entropy alloys spontaneously develop two distinct forms of SRO. Remarkably, the two types of SROs, which carry different energies, distinct degrees of local ordering, and dissimilar electronic structures, are found to co-exist in a wide range of compositions of SiGeSn alloys. The co-existence of two SROs is rationalized through their virtual degeneracy of thermodynamic stability, due to the subtle balance in the change of enthalpy and configurational entropy upon the transformation between the two SROs.  Such co-existence of SROs thus suggests an inherent structural heterogeneity, a diffuse electronic structure, and a new route for band engineering in SiGeSn MEA. More generally, our finding indicates the possible ubiquity of the co-existence of multiple forms of SRO in a broad range of MEAs and HEAs, which has profound implications on their diverse physical properties.             

\end{abstract}

\pagebreak

\section*{Introduction}
Alloys composed of elements with distinct physical and chemical nature often develop a short-range chemical order (SRO), where certain types of elements either attract or repel each other, leading to a distribution of atoms apart from a perfect randomness. Although SRO has been long conceived to exist in alloys \cite{10.1103/physrev.77.669}, the explicit roles of SRO have been identified only recently, including, for example, modulating mechanical properties in metallic alloys \cite{Zhang:2020kr,10.1016/j.jmst.2020.06.018}, enhancing thermoelectric figure of merit in semiconducting alloys \cite{Jiang:2021el,Roychowdhury:2021ut}, and controlling ion transport in oxide alloys \cite{Ji:2019ft}. In particular, very recent breakthrough in atomic-resolution imaging has led to direct observations of SRO in metallic alloys \cite{10.1126/sciadv.aax2799,Zhang:2020kr,Chen:2021bw}, confirming prior theoretical predictions \cite{10.1016/j.actamat.2015.08.015,10.1103/physrevb.91.224204,Ding:2018im}, and more importantly, explicitly demonstrating the impact of SRO on alloys' properties. 

The development of ordering in random alloys can be rationalized through the energy gain when certain types of neighbors form more (or less) frequently than in a truly random distribution. As a result, an alloy can effectively lower its enthalpy of mixing $\Delta H_{\text{mix}}$ through adopting an ordering. However, the development of ordering also leads to a deviation from a random distribution, thus lowering configuration entropy of mixing $\Delta S_{\text{mix}}$. Therefore, depending on the magnitude of the change in $\Delta H_{\text{mix}}$ and $\Delta S_{\text{mix}}$, three possible scenarios may occur: First, when enthalpy gain due to ordering $|\delta \Delta H_{\text{mix}}|$ significantly overweights entropy loss $|-T\delta \Delta S_{\text{mix}}|$, alloy strongly favors ordering, and possibly even develops a long-range order (LRO), {\em i.e.,}  formation of compound or intermetallics; Second, when enthalpy grain cannot compensate entropy loss, a random distribution is thermodynamically preferred; Third, when enthalpy gain becomes greater than entropy loss but not by too much, alloy develops a thermodynamic driving force sufficient to favor a local ordering but not enough for a LRO. This is where SRO emerges. 

Therefore, the occurrence of SRO in an alloy strongly depends on the free energy landscape in alloy's configurational space. For such, some binary alloys can already develop an SRO even though their configurational space is relatively small \cite{10.1126/sciadv.aax2799,Cao:2020hs,10.1103/physrevmaterials.5.104606}. A ternary alloy, particularly a medium-entropy alloy (MEA), has a significantly larger configurational space, and the corresponding free energy landscape becomes much more complex. Accordingly, SRO is expected to be more likely in MEA and high-entropy alloy (HEA). More importantly, the increasing complexity of the free energy landscape may also well indicate SRO can bear more complex forms \cite{10.1038/s41467-019-11464-7}. Indeed, very recent simulation study \cite{10.1038/s41467-021-25264-5} showed the existence of SRO in CoCuFeNiPd HEA can lead to a pseudo-composite microstructure upon mechanical deformation, implying the complexity of SRO. Nevertheless, regardless of the conjecture, it is unclear whether SRO itself can explicitly exhibit structurally distinct forms, which, despite their unique local environment, may co-exist in one alloy.      

Here we investigate the SRO behaviors in SiGeSn MEA alloys that have attracted a substantial interest as potential candidates for mid-infrared photonics owing to their tunable band gap, low-cost, and Si-compatibility \cite{Soref:1991ce,Gencarelli:2013go,Wirths:2016ku,Ghetmiri:2014hh, Wirths:2015dh,Stange:2016ip,Reboud:2017cr,AlKabi:2016de, Margetis:2017fs,Li:2018ka,Kurosawa:2015ba,Tolle:2006js,10.1063/5.0063179}. Although group IV alloys have been commonly conceived as random solid solutions, our recent studies showed both GeSn \cite{Cao:2020hs} and SiSn \cite{10.1103/physrevmaterials.5.104606} binary alloys exhibit substantial SRO behaviors. Upon the mixing of all three elements, we find, through an extensive {\em ab initio}-based Monte Carlo (MC) sampling study, that the SiGeSn MEA exhibits two distinct types of SROs with unique structural signatures and clearly distinguished electronic properties. Remarkably, our study shows the two types of SROs spontaneous occur in MC trajectories, indicative of their co-existence in SiGeSn alloys. Through explicitly computing their enthalpy gain and entropy loss with respect to a random alloy, we confirm the virtual degenerate stability of the two types of SROs.  

\section*{Results}

To explore SRO of an alloy, we employ Metropolis Monte Carlo method combined with Density Functional Theory (MC/DFT) (see Method), which samples alloy's configurational space with high-fidelity. An important point that should be taken into consideration is the significant configurational space of a MEA alloy. For example, the total number of configurations for a 64-atoms cell of the binary Ge$_{0.75}$Sn$_{0.25}$ alloy is on the order of $10^{14}$ (including symmetry-equivalent structures), while that of Si$_{0.25}$Ge$_{0.5}$Sn$_{0.25}$ becomes $10^{27}$. Therefore, to ensure sufficient sampling, we carry out 5$\sim$8 independent MC sampling studies, each consisting of 15,000$\sim$ 20,000 MC moves, from different initial random structures for each alloy composition. This leads to a level of sampling $\sim$ 2,000 moves per atom, and a total number of over 1 million DFT calculations for the entire investigated composition domain.     

\subsection*{Energy landscape and structures of two types of SROs} 
\begin{figure}[H]
\begin{center}
\includegraphics[width=1\linewidth]{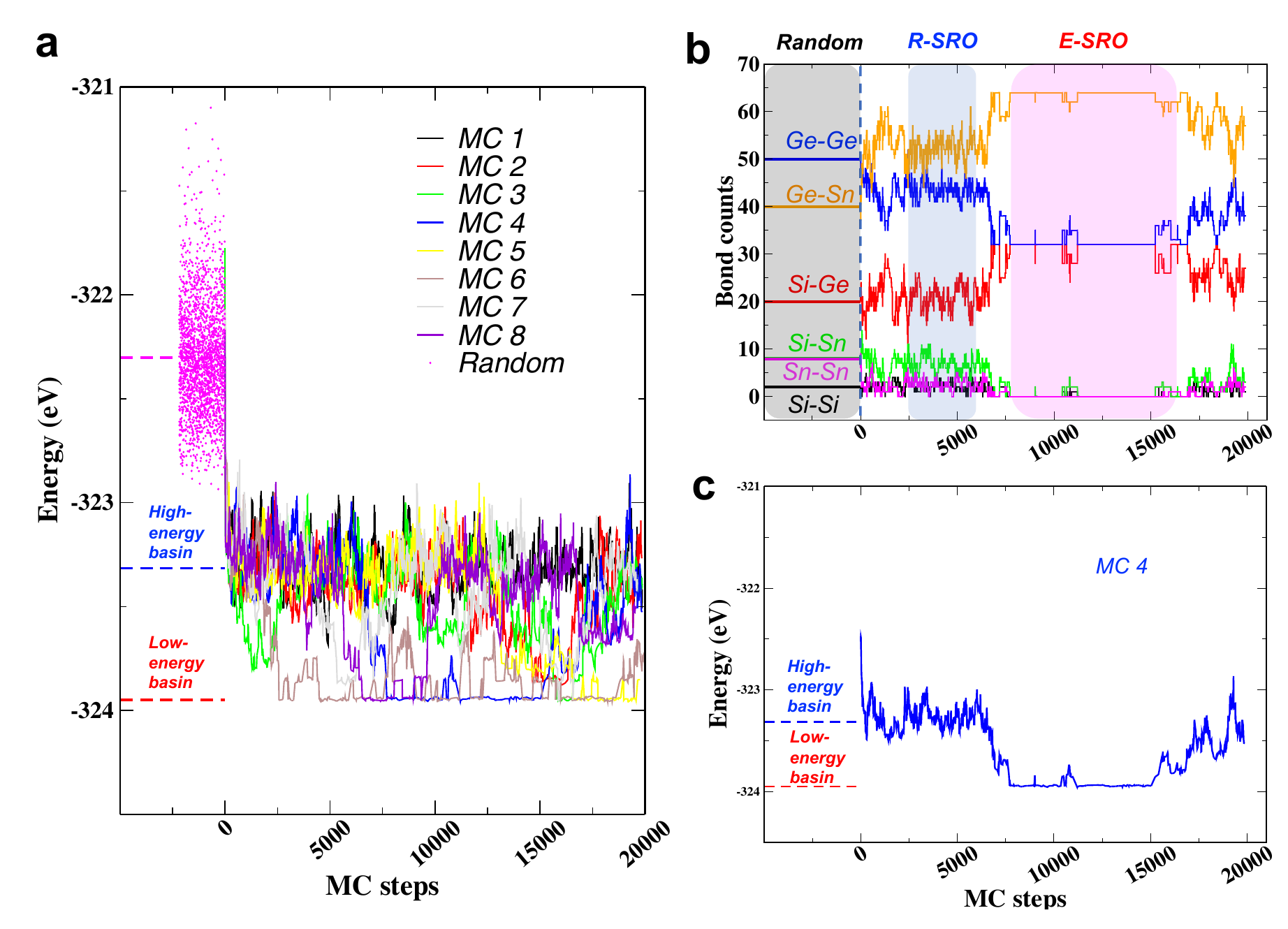}
\caption{\label{energy-structure} Variation of energy and structure due to different types of SROs in Si$_{0.125}$Ge$_{0.625}$Sn$_{0.25}$. ({\bf a}) The overlay of eight independent MC/DFT trajectories shows the total energy fluctuates around two well-defined energy levels: the high-energy basin (blue dashed line) and the low-energy basin (red dashed line) that are separated by $\sim$0.6 eV. Both are well below the average energy (magenta dash line) from random sampling (magenta dots). For clarity, the energy variation of MC 4 is shown separately in ({\bf c}). In accordance, ({\bf b}) displays the corresponding variation of bond counts for each type of nearest neighbors. For a truly random distribution, the number of each type of bond with a 64-atom cell (denoted by the horizontal solid lines within the gray-shaded area) is determined only by the overall alloy composition. R-SRO leads to a different combination of six types of bonds (blue-shaded area), yielding a deviation from a random distribution, whereas E-SRO (magenta-shaded area) enhances such deviation.}
\end{center}
\end{figure}

The large-scale MC/DFT study identifies the complex energy landscape of SiGeSn MEA alloy. As shown in Fig. \ref{energy-structure}{\bf a}, the energy of Si$_{0.125}$Ge$_{0.625}$Sn$_{0.25}$ is found to rapidly decrease at the initial stage for all the eight MC samplings. Such decrease in energy is usually indicative of the existence of an SRO, because rearranging atomic configurations leads to optimized structures which are energetically preferred over a random alloy structure. Indeed, similar behaviors were shown in the alloys previously identified to contain SRO \cite{10.1016/j.actamat.2015.08.015,10.1103/physrevb.91.224204,Ding:2018im,Cao:2020hs,10.1103/physrevmaterials.5.104606,10.1038/s41467-021-25264-5}. Intriguingly, the MC sampling is then found to converge into two, rather than one, energy levels, which has not been reported in prior studies. The two energy levels, denoted as high-energy basin and low-energy basin, respectively, are both well below the average energy of a random alloy, but are separated from each other by about 0.6 eV, as shown in Fig \ref{energy-structure}{\bf a}\&{\bf c}.  We note that this energy difference is much greater than the typical magnitude of energy variation due to thermal fluctuation at room temperature.

The large energy difference between the two energy basins may thus well indicate a distinction in the structures associated with the two energy basins. To examine the structural origin, we examine the statistics of all the six types of first nearest neighbors in Si$_{0.125}$Ge$_{0.625}$Sn$_{0.25}$, namely, Si-Si, Si-Ge, Si-Sn, Ge-Ge, Ge-Sn, and Sn-Sn. Fig. \ref{energy-structure}{\bf b} shows the variation of the instantaneous count of each type of pairs closely follows the variation of energy along the MC trajectory, thus confirming the strong correlation between energy and structure. Importantly, the two energy basins are found to correspond to two distinct combinations of bond counts, respectively, indicating two different types of SROs in SiGeSn alloy. 

\begin{figure}[h]
  \includegraphics[width=0.55\linewidth]{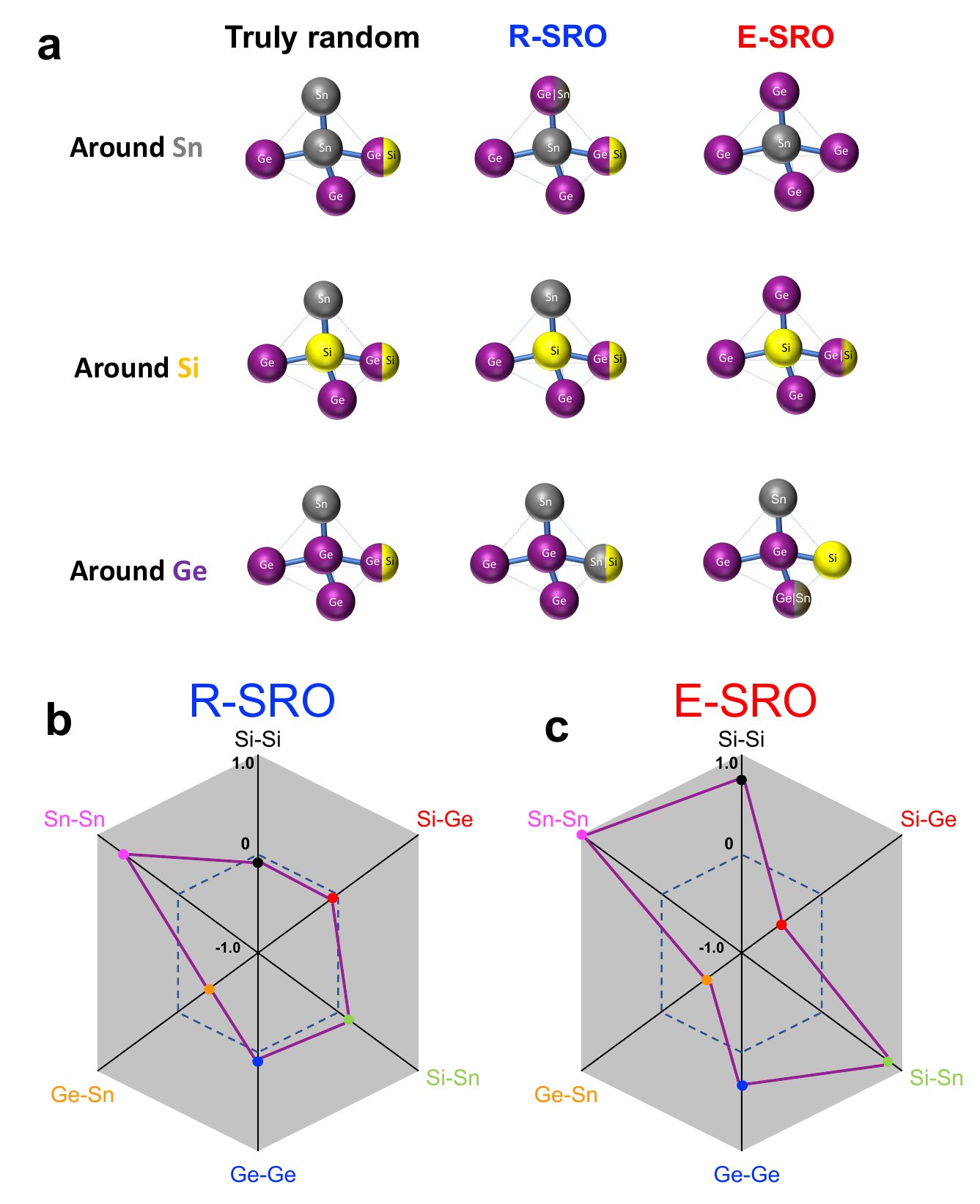}
\caption{\label{motif} Structural signatures of R-SRO and E-SRO in Si$_{0.125}$Ge$_{0.625}$Sn$_{0.25}$ ({\bf a}) The averaged local tetrahedral structural motif around each alloying element when the alloy adopts a random distribution, R-SRO and E-SRO. ({\bf b}) and ({\bf c}) show the SRO polygons constructed from the averaged SRO parameter $\alpha_{ij}^1$ for each type of nearest neighbor in R-SRO and E-SRO, respectively. The averaged SRO parameter $\alpha_{ij}^1$ for the pair $i-j$ is depicted as a dot on the axis labeled as $i-j$, with its location determined by the value of $\alpha_{ij}^1$. An SRO polygon (purple solid lines) is obtained through connecting the six dots. The dashed-hexagon represents a random distribution where $\alpha_{ij}^1=0$. It is visibly clear that only Ge-Sn and Sn-Sn in R-SRO display substantial deviations from the random distribution, whereas the six types of pairs in E-SRO all show strong non-random distributions.}
\end{figure}

To understand the nature of the two types of SROs, we sample in both energy basins and compute the corresponding average tetrahedral structures around a center atom and the SRO parameter $\alpha_{ij}^m$ for the pair between element $i$ and $j$ (see Methods for its definition). Fig. \ref{motif}{\bf a} compares the local structural motifs of a truly random alloy with those identified in the two types of SROs. For a truly random Si$_{0.125}$Ge$_{0.625}$Sn$_{0.25}$ alloy, each center atom, regardless of its type, is surrounded by 0.5 Si atoms, 2.5 Ge atoms, and 1 Sn atom on average, as required by the overall alloy composition. When the configuration energy falls within the high-energy basin, a major change is found to occur around Sn center atoms: a significant portion of Sn-Sn bonds are replaced by Ge-Sn bonds, leading to a substantially lower-than-random Sn-Sn first coordination number. In comparison, the local configuration around Si atom remains virtually intact. Ge atoms, on the other hand, have a portion of their nearest-neighbor Ge atoms replaced by Sn atoms, as a result of the depletion of Sn-Sn bonds and the conservation of the total number of bonds. The resulted SRO parameter for each pair is shown in Fig. \ref{motif}{\bf b}, which clearly indicates SRO occurs mainly through the Sn-Sn and Sn-Ge distribution. We note that this overall structural signature is virtually identical to that identified in Ge$_{0.75}$Sn$_{0.25}$, which is attributed to the Sn-Sn repulsive interaction \cite{Cao:2020hs}. Therefore we denote this type of SRO as regular-SRO (R-SRO). 

In comparison, when the energy falls within the low-energy basin, all the structural motifs are found to undergo changes with respect to a random alloy. The most prominent change occurs in the first coordination shell of Sn atoms, where all the Si and Sn atoms are found to be replaced by Ge atoms. As a result, each Sn atom is surrounded by 4 Ge atoms. Since this change leads to a local Sn structure more ordered than that in R-SRO, we refer it as enhanced-SRO (E-SRO). The elimination of Sn-Sn and Sn-Si bonds then yields the corresponding changes in Si and Ge local structures. For Si, the complete depletion of Sn atoms from its first coordination shell means the bond conservation can only be achieved through Si neighboring with more Ge and/or Si atoms. Our analysis shows that in the case of Si$_{0.125}$Ge$_{0.625}$Sn$_{0.25}$, this change is mainly achieved through the augmentation of Si-Ge bonds. On the other hand, the increased number of both Sn-Ge and Si-Ge bonds must be balanced by a depletion of Ge-Ge bonds, as required by the conservation of total number of bonds around Ge. Correspondingly, the Ge-Ge coordination number in E-SRO is found to be lower than that of R-SRO. Therefore, all six types of pairs in E-SRO are found to display substantial deviations from the random distribution, with the most significant deviations occurring in Sn-Sn, Si-Sn, and Si-Si pairs, which lead to a severely distorted SRO polygon, as shown in Fig. \ref{motif}{\bf c}.    

Therefore, the overall structural signature of E-SRO in Si$_{0.125}$Ge$_{0.625}$Sn$_{0.25}$ can be characterized through the strong repulsive interactions of Sn-Sn and Sn-Si nearest neighbors. The Sn-Sn repulsion has been recently identified in both GeSn and SiSn alloy, which was found to yield a substantially low Sn-Sn coordination number in both alloys \cite{Cao:2020hs,10.1103/physrevmaterials.5.104606}. Our current study suggests that combining GeSn and GeSn in a ternary alloy may facilitate a further depletion of Sn-Sn bonds, leading to a nearly complete elimination of Sn-Sn nearest neighbors in E-SRO. On the other hand, the strong repulsive Si-Sn interaction in E-SRO of SiGeSn may appear to be at odd with the SRO identified in SiSn alloy \cite{10.1103/physrevmaterials.5.104606} at first glance, but we note that such interaction in fact further reduces the number of Sn-Si-Sn local motif, which is the major unfavorable configuration identified in SiSn \cite{10.1103/physrevmaterials.5.104606}. Therefore, the E-SRO optimizes the structure of SiGeSn alloy through reducing all the major unfavorable structures collectively in both GeSn and SiSn sub alloys, which consequently enables a further decrease of configurational energy with respect to R-SRO. It is noted that the structural signatures of both types of SRO are not specific to Si$_{0.125}$Ge$_{0.625}$Sn$_{0.25}$, but rather general in a wide range of compositions, as discussed below and shown in the Supporting Information Fig. S1.  To this end, we also note that although an explicit experimental characterization of SRO has proved challenging \cite{10.1038/ncomms6501,Zhang:2017bw,Zhang:2020kr}, there do exist evidences suggesting the existence of SRO in SiGeSn alloy. For example, both atom probe tomography \cite{Mukherjee:2017is} and extended X-ray absorption fine structure (EXAFS) \cite{Shimura:2017gb} studies indicated a repulsive interaction between Si and Sn in SiGeSn alloy, consistent with our finding in E-SRO. In particular, the EXAFS study \cite{Shimura:2017gb} further showed that Sn atoms prefer to be located at the second nearest neighbor of Sn through a Si-Ge-Sn local bonding structure in SiGeSn. Consistently, our analysis clearly demonstrates a significant preference of the Si-Ge-Sn configuration (Fig. \ref{motif}{\bf a} and Supporting Information Fig. S2) in E-SRO over that in random alloy and R-SRO. 

\subsection*{Thermodynamic stability and composition dependence of SROs}

\begin{figure}[t]
  \includegraphics[width=1\linewidth]{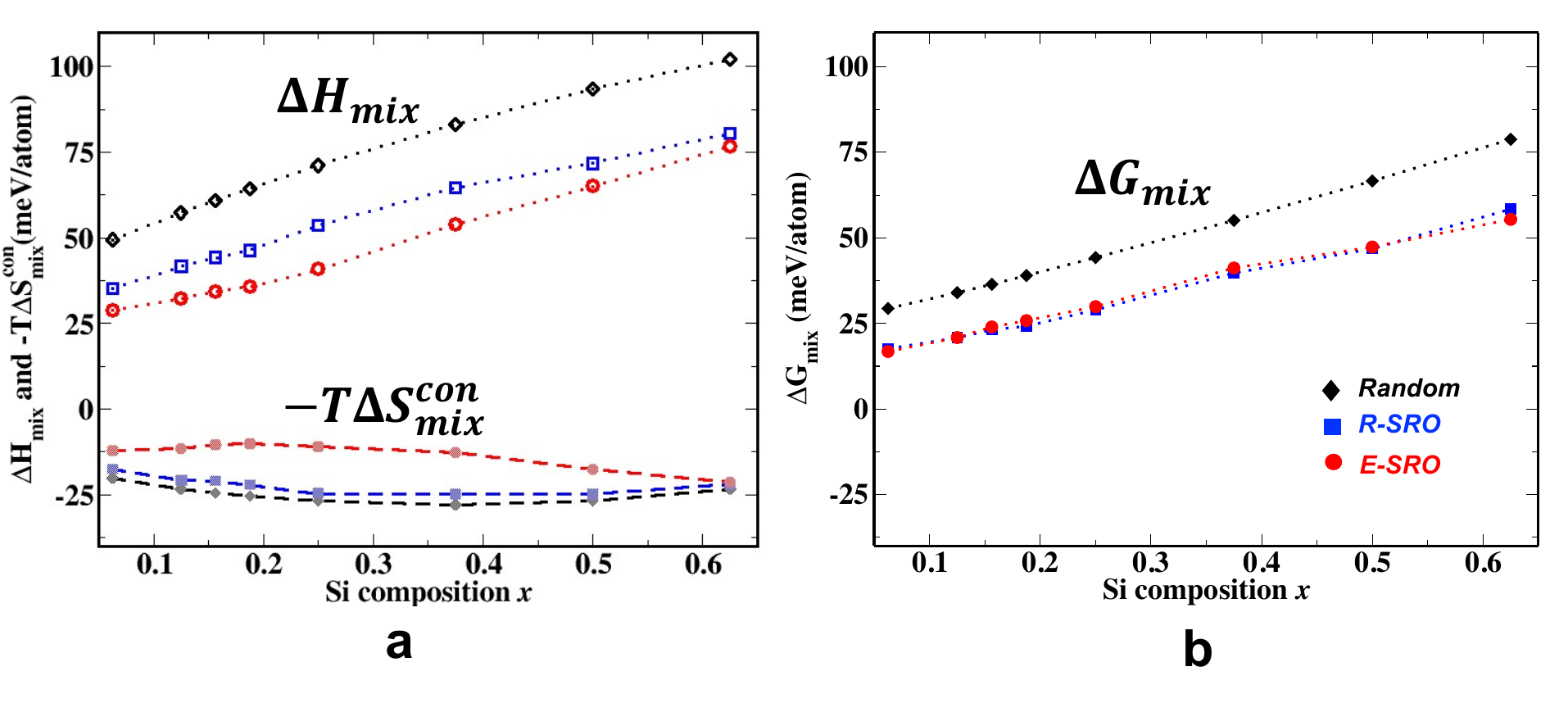}
\caption{\label{TD} Thermodynamic analysis of Si$_{x}$Ge$_{0.75-x}$Sn$_{0.25}$ alloy at room temperature. ({\bf a}) Variation of the calculated enthalpy of mixing $\Delta H_{\text{mix}}$ (open symbols on the top) and the free energy contribution from the configuration entropy of mixing $-T\Delta S_{\text{mix}}^{\text{con}}$ (shaded symbols at the bottom) with Si composition $x$ for a random (black), R-SRO (blue) and E-RSO (red). ({\bf b}) Variation of the calculated total free energy of mixing $\Delta G_{\text{mix}}=\Delta H_{\text{mix}}-T\Delta S_{\text{mix}}^{\text{con}}$ with $x$. R-SRO and E-SRO are nearly degenerate in their $\Delta G_{\text{mix}}$.}
\end{figure}

An intriguing manifestation of the sampling study is the spontaneous occurrence of both R-SRO and E-SRO, albeit E-SRO clearly being energetically more favorable than R-SRO. In fact, Fig. \ref{energy-structure}{\bf a}\&{\bf c} shows the configuration energy fluctuates between the two energy basins in the same trajectories, despite the large difference (0.6 eV) between the two basins. This behavior indicates that both structures contribute to the ensemble average of alloy's properties, and more importantly, that R-SRO and E-SRO may have comparable statistical weights towards the ensemble average. The latter can be likely because, although E-SRO is lower in energy, the further ordering of atoms leads to a loss of configurational entropy, which offsets the enthalpy gain. To explicitly examine the thermodynamic stability of SiGeSn alloy with different SROs, we compute the Gibbs free energy of mixing $\Delta G_{\text{mix}}=\Delta H_{\text{mix}}-T\Delta S_{\text{mix}}^{\text{con}}$ for random alloy, R-SRO, and E-SRO. The mixing enthalpy $\Delta H_{\text{mix}}$ can be readily calculated through averaging the energy in the corresponding parts of the trajectories from MC sampling (for R-SRO and E-SRO) or from random sampling (for random alloy), while the mixing configurational entropy $\Delta S_{\text{mix}}^{\text{con}}$ is estimated based on the extension to quasi-chemical approximation \cite{10.1103/physrevb.36.4279} (see Method for details). As shown in Fig. \ref{TD}{\bf a}, the random Si$_{0.125}$Ge$_{0.625}$Sn$_{0.25}$ alloy yields an $\Delta H_{\text{mix}}$ of 57.3 meV/atom, which is reduced to 41.7 meV/atom and 32.4 meV/atom for the R-SRO and E-SRO, respectively. On the other hand, the random alloy leads to the highest configurational entropy of mixing, which, at the room temperature, corresponds to the lowest $-T\Delta S_{\text{mix}}^{\text{con}}$ of -23.3 meV/ atom. Expectedly, SRO is found to raise $-T\Delta S_{\text{mix}}^{\text{con}}$, through a moderate increase to -20.8 meV/atom via R-SRO and a more significant increase to -11.5 meV/atom via E-SRO.   Remarkably, the calculated $\Delta G_{\text{mix}}$ are found to be nearly identical for R-SRO (20.9 meV/atom) and E-SRO (20.8 meV/atom), which are significantly lower than that of random alloy (34 meV/atom). This virtual degeneracy of $\Delta G_{\text{mix}}$ thus explains the spontaneous co-occurrence of R-SRO and E-SRO in Si$_{0.125}$Ge$_{0.625}$Sn$_{0.25}$ from MC sampling.  

The analysis then leads to a question as to whether the co-existence is of a general relevance at other compositions. Obviously, an exhaustive search over the entire compositional space of SiGeSn can be computationally demanding. In view of the active role of configurational entropy, we address this question by limiting our investigation near the center of the compositional space of the ternary alloy. For each composition, we carry out 5$\sim$8 independent MC/DFT  samplings, with a total number of MC steps comparable to that of Si$_{0.125}$Ge$_{0.625}$Sn$_{0.25}$. The sampling study leads to a total of over 1 million DFT calculations (see Supporting Information Fig. S4-S14). Through this massive sampling study, we identify two types of compositions: namely, the compositions that clearly demonstrate the co-existence of both types of SROs, and those that display the sign of coexistence but have not explicitly exhibited the behavior within the length of the sampling. The explored compositions are indicated in the Si-Ge-Sn pseudo-phase diagram, as shown in the Fig S3. From the compositions that show a clear coexistence, we can obtain a meaningful average of energy and structure within the corresponding high- and low-energy basins in the MC trajectories, which enables a similar thermodynamic analysis to compute $\Delta G_{\text{mix}}$ for R-SRO, E-SRO, and random distribution. The calculation indeed shows the co-existence of two types of SROs is generally associated with the virtual degeneracy of their $\Delta G_{\text{mix}}$, as shown in Fig. \ref{TD}{\bf b}.

Finally, it is noted that other contributions to $\Delta G_{\text{mix}}$ also include the vibrational Helmholtz free energy $F_{\text{mix}}^{\text{vib}}$ and electronic free energy $F_{\text{mix}}^{\text{el}}$, which are not included in MC sampling. Although electronic contribution is typical small at room temperature, vibrational contribution may not be always negligible \cite{Manzoor:2018er,10.1038/s41467-021-25979-5}. To understand the role of vibrational free energy on the relative stability of two types of SROs, we randomly choose representative configurations from MC trajectories and compute $F_{\text{mix}}^{\text{vib}}$ by integrating the calculated phonon density of states through harmonic approximation (see Method). As shown in Supporting Information Table S1, the difference in $F_{\text{mix}}^{\text{vib}}$ between R-SRO and E-SRO is found to be less than 0.2 meV/atom, indicating that $F_{\text{mix}}^{\text{vib}}$ is nearly independent of the degree of ordering, thus dose not affect the relative stability of SRO structures.  

\subsection*{Impact on electronic band structures}

\begin{figure}[H]
  \includegraphics[width=0.5\linewidth]{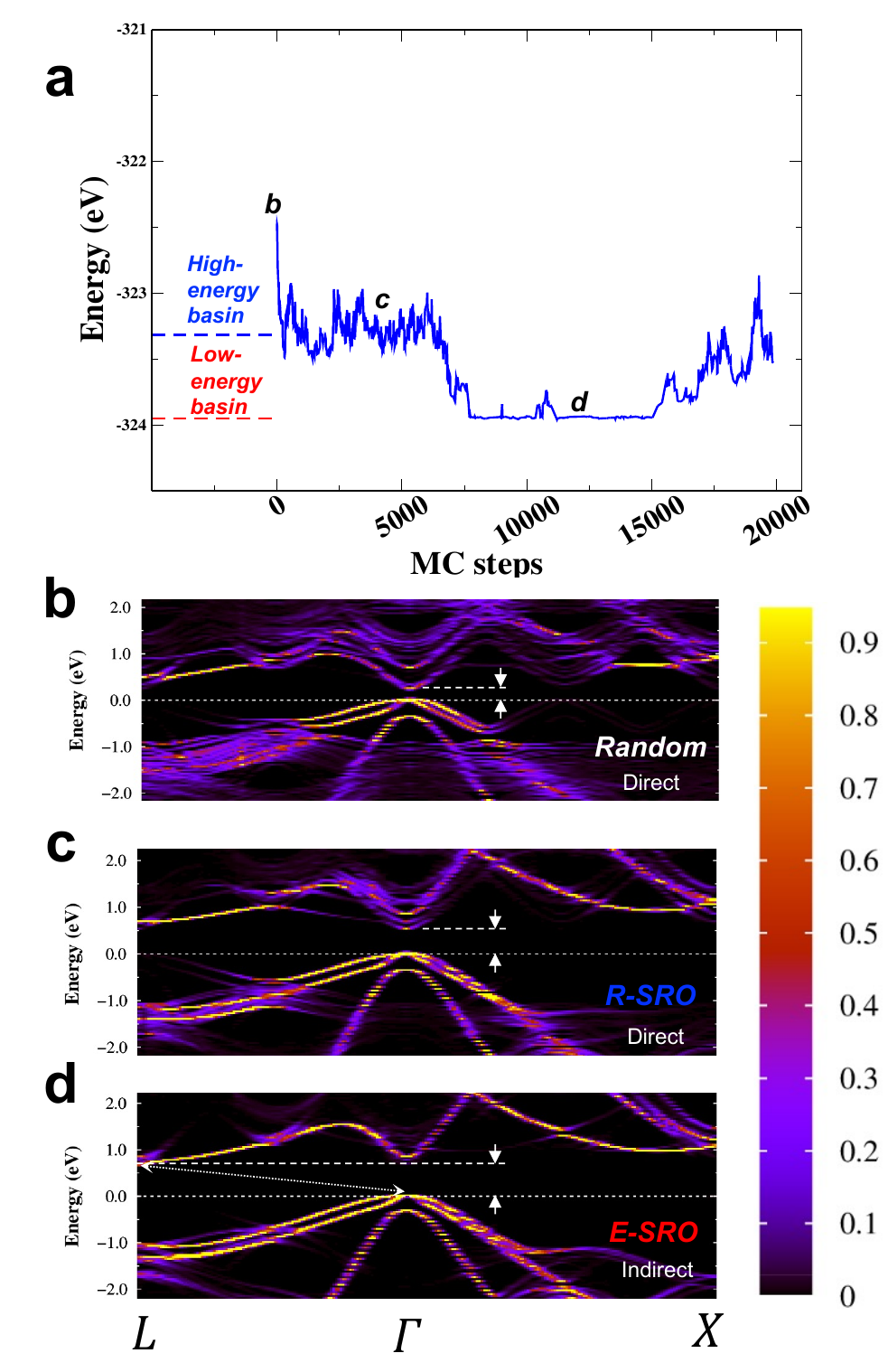}
\caption{\label{BS} Variation of the calculated electronic band structure with SRO in Si$_{0.125}$Ge$_{0.625}$Sn$_{0.25}$. ({\bf a}) A random distribution leads to a direct band gap at $\Gamma$ of 0.25 eV. With R-SRO ({\bf b}), the direct band is found to increase to 0.5 eV. ({\bf c}) Further ordering through E-SRO yields an indirect band gap of $\sim$0.75 eV, defined by a transition between $\Gamma$ and $L$. Band structures of alloy are calculated by unfolding the wavefunction obtained from a 128-atom cell (see Methods) to recover the Bloch character of electronic eigenstates in the Brillouin zone of a diamond cubic lattice. The corresponding Bloch spectral weight is color coded in the bar on the right.}
\end{figure}

The occurrence of SRO has been shown to substantially affect the band gaps in both GeSn \cite{Cao:2020hs} and SiSn \cite{10.1103/physrevmaterials.5.104606} alloys. In light of the greater variation in alloy's structures due to different SRO in SiGeSn, a significant change in electronic structures is expected. To explore this, we compare the calculated band structures of Si$_{0.125}$Ge$_{0.625}$Sn$_{0.25}$ alloy with a random distribution, R-SRO, and E-SRO. As shown in Fig. \ref{BS}, the calculated band structures display salient distinctions among the three types of structures. A random alloy is found to exhibit a direct band gap at $\Gamma$, with the magnitude of  $\sim$0.25 eV. With R-SRO, the gap is found to increase to $\sim$0.5 eV, while maintaining the direct character. Remarkably, when alloy adopts E-SRO, the direct gap at $\Gamma$ sees a further significant increase so that the conduction band minimum at $\Gamma$ is slightly above that at $L$, thus turning the band gap into an indirect type. The corresponding indirect band gap is $\sim$0.75 eV. This significant change in band gaps and band characters thus highlights the substantial role of alloy structure, particularly the degree of SRO, plays in the optoelectronic properties of group IV alloys. 

\section*{Discussion}

The co-occurrence of R-SRO and E-SRO observed in our modeling study, along with the analysis showing their nearly degenerate thermodynamic stability, thus may well indicate a co-existence of both types of SROs in real SiGeSn MEA. Since a dynamic transition between the two SROs requires re-arrangement of atoms in the lattice, which can be kinetically prohibited at room temperature, such co-existence is more likely to bear a static form. One possible scenario for static co-existence is that SiGeSn MEAs are composed of iso-compositional domains with different degrees of SRO. Although the sizes of these domains are unclear, they are expected to be significantly smaller than a typical grain, and possibly around nanoscale, because the interface separating different SRO domains does not involve energetically unfavorable structures such as crystalline defects, thus the interfacial free energy is expected to be low. Consequently, SiGeSn alloys at equilibrium are expected to be structurally and spatially inhomogeneous at the {\em microscopic} level, albeit being macroscopically homogeneous and isotropic. 

Regardless of its spatial distribution, the identified co-existence of two types of SROs has significant implications on the optoelectronic properties of SiGeSn alloys. First, the prediction of the key optoelectronic properties of SiGeSn alloy, {\em e.g.}, band gap, type of band gap, critical compositions for indirect-to-direct transition, must take into account the significant role of SRO. Given the large difference in the calculated band gap ($\sim$0.5 eV) between a random and E-SRO model for Si$_{0.125}$Ge$_{0.625}$Sn$_{0.25}$, it is highly likely that a prediction based on a random structure model may substantially underestimate the band gap. This has already been shown in GeSn \cite{Cao:2020hs} and SiSn \cite{10.1103/physrevmaterials.5.104606} alloys, but is expected to be even more significant in ternary alloys. In particular, since the currently projected line of composition for indirect-to-direct transition in SiGeSn \cite{10.1063/5.0043511} (see Supporting Information Fig. S3) significantly overlaps the composition range where the co-existence of two SROs is clearly identified, a subsequent revisit to the current model is deemed necessary. Second, the measured optoelectronic properties of SiGeSn alloys reflect an ensemble average over all the contributions from the different structures with various degrees of SRO. Given the large variation in the electronic structures of SRO with different degrees that can co-exist, an averaged property {\em alone}, {\em e.g.}, band gap, may not be sufficient to characterize alloy's optoelectronic properties. For example, SiGeSn alloys may exhibit smeared band edges, analogous to band tails in heavily doped semiconductors \cite{10.1103/revmodphys.64.755}, instead of a rather abrupt, step-like onset of absorption coefficient as in a compound semiconductor. Such optoelectronic properties may affect the performance of the alloys, depending on exact types of potential applications. To this end, we note that the projected compositions of SiGeSn alloy for the proposed key mid-infrared applications (see Ref. \cite{10.1103/physrevmaterials.5.104606} and Supporting Information Fig. S3) significantly overlaps the compositional domain where the co-existence of SROs is identified. Third, the significant variation in electronic structure, which is solely due to the change in SRO, may also present an opportunity for band engineering. A critical limitation in combining two dissimilar band structures to create a heterojunction, {\em e.g.}, quantum well, is the lattice mismatch between two different materials, which stands as one of the key challenges in epitaxial growth \cite{10.1146/annurev-matsci-070909-104448}. In this regard, a heterojunction composed of SiGeSn alloys with identical composition but different SROs, for example, R-SRO and E-SRO, can be advantageous, because lattice constant of alloys only displays a very weak dependence on ordering (see Supporting Information Table S2). Therefore, SRO presents as an additional tuning capability for band engineering, and in combining with the large compositional space of SiGeSn alloy, a variety of types of heterojunction may be formed on demand.       

Lastly, and perhaps even more importantly, it is expected that the observed phenomenon can exist generally in many other alloys. The co-existence of the two SROs is essentially due to the subtle balance between the enthalpy gain and the entropy loss upon an enhanced ordering of alloy configurations. Certainly, such balance cannot be always expected in any alloy, thus the discovery of the co-existence of two SROs in SiGeSn MEA can be a fortunate case in this sense. However, we note that as the number of alloying elements increases, the configurational space of alloy increases combinatorially. As a consequence, both the energy landscape and free energy landscape become inevitably more complex. Therefore it is increasingly likely that there exist multiple local free energy minima with comparable stability in the SRO-parameter space. If this were true, multiple forms of SRO could emerge at equilibrium. These forms of SRO can be reflected by different combinations of various neighbors within the first few atomic shells that are nearly equally favorable over a random distribution. As a consequence, these alloys can be intrinsically heterogeneous in nanoscale, which has profound implications on the mechanical, optoelectronic, and thermal transport properties. In this regard, we note that MEA and HEA are very probable alloy systems to potentially exhibit this type of behavior. In fact, some of the very recent studies have already provided the sign for this. For example, a theoretical study on CoCuFeNiPd HEA \cite{10.1038/s41467-021-25264-5} has demonstrated that SRO induces a composite structure consisting of three different types of clusters. Experimental characterization work has also observed concentration waves at nanoscale in both CrFeCoNiPd HEA \cite{10.1038/s41586-019-1617-1} and VCoNi MEA \cite{Chen:2021bw}. Therefore, we expect more discoveries to follow as SRO in MEA and HEA is being intensively investigated.      

\section{Methods}		
\subsection{Monte Carlo (MC) Sampling}
The Metropolis Monte Carlo (MC) method \cite{Metropolis:1953in} is used to obtain the ensemble averages accounting for SRO,  where the acceptance probability for a new configuration $j$ generated from a trial move in configuration \(i\) equals to $\min$\(\{1,\exp(-(E_j-E_i)/k_BT)\}\), where \(E_i\) and \(E_j\) are the total energies of configuration \(i\) and \(j\), respectively, \(k_B\) is the Bolzmann constant, and \(T\) is temperature. Each trial move involves randomly selecting and swapping a pair of two atoms of different species to create a new configuration $j$, which then undergoes a full relaxation to obtain its energy \(E_j\). 

\subsection{DFT calculations}
Our density functional theory (DFT) calculations were performed with Vienna {\em ab initio} simulation package (VASP) \cite{Kresse:1993ty} based on the projector augmented wave method \cite{Kresse:1999tq,Kresse:1996kg,Kresse:1996vf}. Local density approximation (LDA) \cite{Ceperley:1980gc} was employed for the exchange-correlation functional, which has been shown to yield the best agreement with experiment on pure Ge and Sn for geometry optimization \cite{Eckhardt:2014gz,Polak:2017fh,Haas:2009be,Tran:2009kk} 
The simulation cell containing 64 atoms is obtained by replicating a conventional DC cell containing eight atoms twice along each dimension. We used $2\times2\times2$ Monkhorst-Pack \(k\)-points grid \cite{Monkhorst:1976cv} with a plane-wave cutoff energy of 300 eV. 
The conjugate-gradient algorithm is applied for structural relaxation during each energy calculation, with the convergence criteria of \(10^{-4}\) eV and \(10^{-3}\) eV for electronic and ionic relaxations, respectively. 
To cross-validate the results against size and shape effects, we also carried out MC sampling based on 128-atom cells created by replicating the primitive cell of the DC structure four times along each dimension. In this case only gamma point is used to sample the Brillouin zone for higher computational efficiency. We note that the choice of all these parameters has been carefully examined to ensure the total energy difference between two configurations $i$ and $j$ ($\Delta E = E_j - E_i$), which plays a crucial role in MC sampling, is well converged.

\subsection{SRO parameter}
The SRO parameter $\alpha_{ij}^m$ is defined as 

\begin{equation} \label{eq1}
\alpha_{ij}^m = 1 - \frac{N_{ij}^m}{N_{0,ij}^m},
\end{equation}
where $N_{0,ij}^m$ and $N_{ij}^m$ are the numbers of pairs between element $i$ and $j$ in the $\text{m}^{\text{th}}$ coordination shell for a random distribution and the actual distribution, respectively. By definition, a random distribution yields $\alpha_{ij}^m$ = 0. $0<\alpha_{ij}^m\le 1$ indicates a depletion of $i,j$ pairs, with $\alpha_{ij}^m=1$ meaning a complete depletion, while $\alpha_{ij}^m$ $<$ 0 suggests a preference of $i,j$ pairs. Note that $\alpha_{ij}^m$  is different from but related to the commonly adopted Warren-Cowley SRO parameter \cite{10.1103/physrev.77.669}. 

\subsection{Quasi-chemical approximation}
The configuration entropy of mixing $\Delta S_{\text{mix}}^{\text{con}}$ is estimated based on the extension of quasi-chemical approximation \cite{10.1103/physrevb.36.4279} from binary alloys to multi-component alloys. For ternary system composed of species $A$, $B$, and $C$ with composition of $x_A$, $x_B$ and $x_C$, the configuration entropy per atom can be approximated by

\begin{equation} \label{eq2}
\Delta S_{\text{mix}}^{\text{con}} = -k_B\left[ \sum_{i=A,B,C} x_i\ln x_i + \sum_{j} \left(x_j\ln x_j - x_j\ln x_j^0\right)\right],
\end{equation}
where $k_B$ is the Bolzmann constant, $x_i$ is the concentration of species $i$, $x_j^0$ is the concentration of the micro-cluster $j$ in the random solution model, and $x_j$ is the concentration of the cluster $j$ in the real solution modeled by MC sampling through taking SRO into account. We use triplets as micro-clusters here to account for SRO reflected by both 1NN and 2NN, and it can be shown that there are a total of 18 types of distinguishable micro-clusters in a ternary alloy. 

\subsection{Band structure calculations}

For band structure calculations, we employ the modified Becke-Johnson (mBJ) exchange potential \cite{Tran:2009kk}, as implemented in VASP code, with the $c$-mBJ parameter set to be 1.2, which has been demonstrated to predict the correct band gaps of Si, Ge and $\alpha-$Sn, in good agreement with experimental data and much computationally efficient than hybrid functionals or GW methods \cite{Tran:2009kk,Eckhardt:2014gz,Polak:2017fh}.  We apply the spectral weight approach \cite{Popescu:2010jd,Rubel:2014fv} to unfold the band structures back into the first primitive Brillouin Zone of diamond cubic structure using the code {\em fold2bloch} \cite{Rubel:2014fv}. Since the spectral weight approach as implemented in {\em fold2bloch} \cite{Rubel:2014fv} requires the supercells built on the basis of primitive cells instead of conventional ones, band structure calculations are carried out on 128-atom supercells transformed from the 64-atom supercells generated from the MC/DFT trajectories, with a $3\times3$ integer transformation matrix $[[0, 1, 1], [1, 0, 1], [1, 1, 0]]$. Relativistic effects (spin-orbit coupling) are included in the band structure calculation, which has been demonstrated to be crucial for reproducing the band structures of Ge and $\alpha-$Sn \cite{Eckhardt:2014gz,Polak:2017fh}.

\subsection{Vibrational Helmholtz Free Entropy Calculations}

Vibrational Helmholtz free energy $F_{\text{mix}}^{\text{vib}}$ is calculated through the harmonic approximation \cite{Dove_1993}, using the {\em phonopy} code \cite{10.1016/j.scriptamat.2015.07.021} interfaced with VASP. For SiGeSn alloy, we compute the second-order interatomic force constants using the finite-displacement method with an atomic displacement of 0.01 \AA ~in supercells containing 512 atoms, duplicated $2\times2\times2$ from the 64-atom supercells generated from the MC/DFT trajectories, using the $\Gamma$-point sampling for electronic self-consistent calculations. For pure Si, Ge, and $\alpha$-Sn, we used supercells containing 128 atoms, duplicated $4\times4\times4$ from the optimized primitive unit cells, with $2\times2\times2$ $k$-points for electronic self-consistent calculations. The post-processing of the results was performed with the {\em phonopy} code \cite{10.1016/j.scriptamat.2015.07.021}. The convergence is carefully checked with the $q$-points meshes up to $8\times8\times8$ for SiGeSn alloy and $35\times35\times35$ for pure Si, Ge, and $\alpha$-Sn.

\begin{acknowledgement}

The authors thank S.-Q. (Fisher) Yu for helpful discussion. This material is based on work supported by the Air Force Office of Scientific Research under Award No. FA9550-191-0341. The authors acknowledge the Department of Defense high High Performance Computing Modernization Program for the computing support.

\end{acknowledgement}


\begin{mcitethebibliography}{55}
\providecommand*\natexlab[1]{#1}
\providecommand*\mciteSetBstSublistMode[1]{}
\providecommand*\mciteSetBstMaxWidthForm[2]{}
\providecommand*\mciteBstWouldAddEndPuncttrue
  {\def\EndOfBibitem{\unskip.}}
\providecommand*\mciteBstWouldAddEndPunctfalse
  {\let\EndOfBibitem\relax}
\providecommand*\mciteSetBstMidEndSepPunct[3]{}
\providecommand*\mciteSetBstSublistLabelBeginEnd[3]{}
\providecommand*\EndOfBibitem{}
\mciteSetBstSublistMode{f}
\mciteSetBstMaxWidthForm{subitem}{(\alph{mcitesubitemcount})}
\mciteSetBstSublistLabelBeginEnd
  {\mcitemaxwidthsubitemform\space}
  {\relax}
  {\relax}

\bibitem[Cowley(1950)]{10.1103/physrev.77.669}
Cowley,~J.~M. {An Approximate Theory of Order in Alloys}. \emph{Physical
  Review} \textbf{1950}, \emph{77}, 669--675\relax
\mciteBstWouldAddEndPuncttrue
\mciteSetBstMidEndSepPunct{\mcitedefaultmidpunct}
{\mcitedefaultendpunct}{\mcitedefaultseppunct}\relax
\EndOfBibitem
\bibitem[Zhang \latin{et~al.}(2020)Zhang, Zhao, Ding, Chong, Jia, Ophus, Asta,
  Ritchie, and Minor]{Zhang:2020kr}
Zhang,~R.; Zhao,~S.; Ding,~J.; Chong,~Y.; Jia,~T.; Ophus,~C.; Asta,~M.;
  Ritchie,~R.~O.; Minor,~A.~M. Short-Range Order and Its Impact on the {CrCoNi}
  Medium-Entropy Alloy. \emph{Nature} \textbf{2020}, \emph{581}, 283--287\relax
\mciteBstWouldAddEndPuncttrue
\mciteSetBstMidEndSepPunct{\mcitedefaultmidpunct}
{\mcitedefaultendpunct}{\mcitedefaultseppunct}\relax
\EndOfBibitem
\bibitem[Wu \latin{et~al.}(2021)Wu, Zhang, Yuan, Huang, Wen, Wang, Zhang, Wu,
  Liu, Wang, Jiang, and Lu]{10.1016/j.jmst.2020.06.018}
Wu,~Y.; Zhang,~F.; Yuan,~X.; Huang,~H.; Wen,~X.; Wang,~Y.; Zhang,~M.; Wu,~H.;
  Liu,~X.; Wang,~H.; Jiang,~S.; Lu,~Z. {Short-range ordering and its effects on
  mechanical properties of high-entropy alloys}. \emph{Journal of Materials
  Science \& Technology} \textbf{2021}, \emph{62}, 214 -- 220\relax
\mciteBstWouldAddEndPuncttrue
\mciteSetBstMidEndSepPunct{\mcitedefaultmidpunct}
{\mcitedefaultendpunct}{\mcitedefaultseppunct}\relax
\EndOfBibitem
\bibitem[Jiang \latin{et~al.}(2021)Jiang, Yu, Cui, Liu, Xie, Liao, Zhang,
  Huang, Ning, Jia, Zhu, Bai, Chen, Pennycook, and He]{Jiang:2021el}
Jiang,~B.; Yu,~Y.; Cui,~J.; Liu,~X.; Xie,~L.; Liao,~J.; Zhang,~Q.; Huang,~Y.;
  Ning,~S.; Jia,~B.; Zhu,~B.; Bai,~S.; Chen,~L.; Pennycook,~S.~J.; He,~J.
  {High-entropy-stabilized chalcogenides with high thermoelectric performance}.
  \emph{Science} \textbf{2021}, \emph{371}, 830--834\relax
\mciteBstWouldAddEndPuncttrue
\mciteSetBstMidEndSepPunct{\mcitedefaultmidpunct}
{\mcitedefaultendpunct}{\mcitedefaultseppunct}\relax
\EndOfBibitem
\bibitem[Roychowdhury \latin{et~al.}(2021)Roychowdhury, Ghosh, Arora, Samanta,
  Xie, Singh, Soni, He, Waghmare, and Biswas]{Roychowdhury:2021ut}
Roychowdhury,~S.; Ghosh,~T.; Arora,~R.; Samanta,~M.; Xie,~L.; Singh,~N.~K.;
  Soni,~A.; He,~J.; Waghmare,~U.~V.; Biswas,~K. {Enhanced atomic ordering leads
  to high thermoelectric performance in AgSbTe2}. \emph{Science} \textbf{2021},
  \emph{371}, 722--727\relax
\mciteBstWouldAddEndPuncttrue
\mciteSetBstMidEndSepPunct{\mcitedefaultmidpunct}
{\mcitedefaultendpunct}{\mcitedefaultseppunct}\relax
\EndOfBibitem
\bibitem[Ji \latin{et~al.}(2019)Ji, Urban, Kitchaev, Kwon, Artrith, Ophus,
  Huang, Cai, Shi, Kim, Kim, and Ceder]{Ji:2019ft}
Ji,~H.; Urban,~A.; Kitchaev,~D.~A.; Kwon,~D.-H.; Artrith,~N.; Ophus,~C.;
  Huang,~W.; Cai,~Z.; Shi,~T.; Kim,~J.~C.; Kim,~H.; Ceder,~G. {Hidden
  structural and chemical order controls lithium transport in cation-disordered
  oxides for rechargeable batteries}. \emph{Nat Commun} \textbf{2019},
  1--9\relax
\mciteBstWouldAddEndPuncttrue
\mciteSetBstMidEndSepPunct{\mcitedefaultmidpunct}
{\mcitedefaultendpunct}{\mcitedefaultseppunct}\relax
\EndOfBibitem
\bibitem[Zhang \latin{et~al.}(2019)Zhang, Zhao, Ophus, Deng, Vachhani, Ozdol,
  Traylor, Bustillo, MorrisJr, Chrzan, Asta, and Minor]{10.1126/sciadv.aax2799}
Zhang,~R.; Zhao,~S.; Ophus,~C.; Deng,~Y.; Vachhani,~S.~J.; Ozdol,~B.;
  Traylor,~R.; Bustillo,~K.~C.; MorrisJr,~J.~W.; Chrzan,~D.~C.; Asta,~M.;
  Minor,~A.~M. {Direct imaging of short-range order and its impact on
  deformation in Ti-6Al}. \emph{Science Advances} \textbf{2019}, \relax
\mciteBstWouldAddEndPunctfalse
\mciteSetBstMidEndSepPunct{\mcitedefaultmidpunct}
{}{\mcitedefaultseppunct}\relax
\EndOfBibitem
\bibitem[Chen \latin{et~al.}(2021)Chen, Wang, Cheng, Zhu, Zhou, Jiang, Zhou,
  Xue, Yuan, Zhu, Wu, and Ma]{Chen:2021bw}
Chen,~X.; Wang,~Q.; Cheng,~Z.; Zhu,~M.; Zhou,~H.; Jiang,~P.; Zhou,~L.; Xue,~Q.;
  Yuan,~F.; Zhu,~J.; Wu,~X.; Ma,~E. {Direct observation of chemical short-range
  order in a medium-entropy alloy}. \emph{Nature} \textbf{2021}, \emph{592},
  712--716\relax
\mciteBstWouldAddEndPuncttrue
\mciteSetBstMidEndSepPunct{\mcitedefaultmidpunct}
{\mcitedefaultendpunct}{\mcitedefaultseppunct}\relax
\EndOfBibitem
\bibitem[Tamm \latin{et~al.}(2015)Tamm, Aabloo, Klintenberg, Stocks, and
  Caro]{10.1016/j.actamat.2015.08.015}
Tamm,~A.; Aabloo,~A.; Klintenberg,~M.; Stocks,~M.; Caro,~A. {Atomic-scale
  properties of Ni-based FCC ternary, and quaternary alloys}. \emph{Acta
  Materialia} \textbf{2015}, \emph{99}, 307--312\relax
\mciteBstWouldAddEndPuncttrue
\mciteSetBstMidEndSepPunct{\mcitedefaultmidpunct}
{\mcitedefaultendpunct}{\mcitedefaultseppunct}\relax
\EndOfBibitem
\bibitem[Singh \latin{et~al.}(2015)Singh, Smirnov, and
  Johnson]{10.1103/physrevb.91.224204}
Singh,~P.; Smirnov,~A.~V.; Johnson,~D.~D. {Atomic short-range order and
  incipient long-range order in high-entropy alloys}. \emph{Physical Review B}
  \textbf{2015}, \emph{91}, 224204\relax
\mciteBstWouldAddEndPuncttrue
\mciteSetBstMidEndSepPunct{\mcitedefaultmidpunct}
{\mcitedefaultendpunct}{\mcitedefaultseppunct}\relax
\EndOfBibitem
\bibitem[Ding \latin{et~al.}(2018)Ding, Yu, Asta, and Ritchie]{Ding:2018im}
Ding,~J.; Yu,~Q.; Asta,~M.; Ritchie,~R.~O. Tunable Stacking Fault Energies by
  Tailoring Local Chemical Order in {CrCoNi} Medium-Entropy Alloys. \emph{PNAS}
  \textbf{2018}, \emph{115}, 8919--8924\relax
\mciteBstWouldAddEndPuncttrue
\mciteSetBstMidEndSepPunct{\mcitedefaultmidpunct}
{\mcitedefaultendpunct}{\mcitedefaultseppunct}\relax
\EndOfBibitem
\bibitem[Cao \latin{et~al.}(2020)Cao, Chen, Jin, Liu, and Li]{Cao:2020hs}
Cao,~B.; Chen,~S.; Jin,~X.; Liu,~J.; Li,~T. {Short-Range Order in GeSn Alloy}.
  \emph{ACS Appl. Mater. Interfaces} \textbf{2020}, \emph{12},
  57245--57253\relax
\mciteBstWouldAddEndPuncttrue
\mciteSetBstMidEndSepPunct{\mcitedefaultmidpunct}
{\mcitedefaultendpunct}{\mcitedefaultseppunct}\relax
\EndOfBibitem
\bibitem[Jin \latin{et~al.}(2021)Jin, Chen, and
  Li]{10.1103/physrevmaterials.5.104606}
Jin,~X.; Chen,~S.; Li,~T. {Short-range order in SiSn alloy enriched by
  second-nearest-neighbor repulsion}. \emph{Physical Review Materials}
  \textbf{2021}, \emph{5}, 104606\relax
\mciteBstWouldAddEndPuncttrue
\mciteSetBstMidEndSepPunct{\mcitedefaultmidpunct}
{\mcitedefaultendpunct}{\mcitedefaultseppunct}\relax
\EndOfBibitem
\bibitem[Li \latin{et~al.}(2019)Li, Sheng, and Ma]{10.1038/s41467-019-11464-7}
Li,~Q.-J.; Sheng,~H.; Ma,~E. {Strengthening in multi-principal element alloys
  with local-chemical-order roughened dislocation pathways}. \emph{Nature
  Communications} \textbf{2019}, \emph{10}, 3563\relax
\mciteBstWouldAddEndPuncttrue
\mciteSetBstMidEndSepPunct{\mcitedefaultmidpunct}
{\mcitedefaultendpunct}{\mcitedefaultseppunct}\relax
\EndOfBibitem
\bibitem[Chen \latin{et~al.}(2021)Chen, Aitken, Pattamatta, Wu, Yu, Srolovitz,
  Liaw, and Zhang]{10.1038/s41467-021-25264-5}
Chen,~S.; Aitken,~Z.~H.; Pattamatta,~S.; Wu,~Z.; Yu,~Z.~G.; Srolovitz,~D.~J.;
  Liaw,~P.~K.; Zhang,~Y.-W. {Simultaneously enhancing the ultimate strength and
  ductility of high-entropy alloys via short-range ordering}. \emph{Nature
  Communications} \textbf{2021}, \emph{12}, 4953\relax
\mciteBstWouldAddEndPuncttrue
\mciteSetBstMidEndSepPunct{\mcitedefaultmidpunct}
{\mcitedefaultendpunct}{\mcitedefaultseppunct}\relax
\EndOfBibitem
\bibitem[Soref and Perry(1991)Soref, and Perry]{Soref:1991ce}
Soref,~R.~A.; Perry,~C.~H. Predicted Band Gap of the New Semiconductor
  {SiGeSn}. \emph{Journal of Applied Physics} \textbf{1991}, \emph{69},
  539--541\relax
\mciteBstWouldAddEndPuncttrue
\mciteSetBstMidEndSepPunct{\mcitedefaultmidpunct}
{\mcitedefaultendpunct}{\mcitedefaultseppunct}\relax
\EndOfBibitem
\bibitem[Gencarelli \latin{et~al.}(2013)Gencarelli, Vincent, Demeulemeester,
  Vantomme, Moussa, Franquet, Kumar, Bender, Meersschaut, Vandervorst, Loo,
  Caymax, Temst, and Heyns]{Gencarelli:2013go}
Gencarelli,~F.; Vincent,~B.; Demeulemeester,~J.; Vantomme,~A.; Moussa,~A.;
  Franquet,~A.; Kumar,~A.; Bender,~H.; Meersschaut,~J.; Vandervorst,~W.;
  Loo,~R.; Caymax,~M.; Temst,~K.; Heyns,~M. Crystalline {Properties} and
  {Strain} {Relaxation} {Mechanism} of {CVD} {Grown} {GeSn}. \emph{ECS J. Solid
  State Sci. Technol.} \textbf{2013}, \emph{2}, P134\relax
\mciteBstWouldAddEndPuncttrue
\mciteSetBstMidEndSepPunct{\mcitedefaultmidpunct}
{\mcitedefaultendpunct}{\mcitedefaultseppunct}\relax
\EndOfBibitem
\bibitem[Wirths \latin{et~al.}(2016)Wirths, Buca, and Mantl]{Wirths:2016ku}
Wirths,~S.; Buca,~D.; Mantl,~S. SiÐ{Ge}Ð{Sn} Alloys: {From} Growth to
  Applications. \emph{Progress in Crystal Growth and Characterization of
  Materials} \textbf{2016}, \emph{62}, 1--39\relax
\mciteBstWouldAddEndPuncttrue
\mciteSetBstMidEndSepPunct{\mcitedefaultmidpunct}
{\mcitedefaultendpunct}{\mcitedefaultseppunct}\relax
\EndOfBibitem
\bibitem[Ghetmiri \latin{et~al.}(2014)Ghetmiri, Du, Margetis, Mosleh, Cousar,
  Conley, Domulevicz, Nazzal, Sun, Soref, Tolle, Li, Naseem, and
  Yu]{Ghetmiri:2014hh}
Ghetmiri,~S.~A.; Du,~W.; Margetis,~J.; Mosleh,~A.; Cousar,~L.; Conley,~B.~R.;
  Domulevicz,~L.; Nazzal,~A.; Sun,~G.; Soref,~R.~A.; Tolle,~J.; Li,~B.;
  Naseem,~H.~A.; Yu,~S.-Q. Direct-Bandgap {GeSn} Grown on Silicon with 2230 nm
  Photoluminescence. \emph{Appl. Phys. Lett.} \textbf{2014}, \emph{105},
  151109\relax
\mciteBstWouldAddEndPuncttrue
\mciteSetBstMidEndSepPunct{\mcitedefaultmidpunct}
{\mcitedefaultendpunct}{\mcitedefaultseppunct}\relax
\EndOfBibitem
\bibitem[Wirths \latin{et~al.}(2015)Wirths, Geiger, von~den Driesch, Mussler,
  Stoica, Mantl, Ikonic, Luysberg, Chiussi, Hartmann, Sigg, Faist, Buca, and
  GrŸtzmacher]{Wirths:2015dh}
Wirths,~S.; Geiger,~R.; von~den Driesch,~N.; Mussler,~G.; Stoica,~T.;
  Mantl,~S.; Ikonic,~Z.; Luysberg,~M.; Chiussi,~S.; Hartmann,~J.~M.; Sigg,~H.;
  Faist,~J.; Buca,~D.; GrŸtzmacher,~D. Lasing in Direct-Bandgap {GeSn} Alloy
  Grown on {Si}. \emph{Nature Photonics} \textbf{2015}, \emph{9}, 88--92\relax
\mciteBstWouldAddEndPuncttrue
\mciteSetBstMidEndSepPunct{\mcitedefaultmidpunct}
{\mcitedefaultendpunct}{\mcitedefaultseppunct}\relax
\EndOfBibitem
\bibitem[Stange \latin{et~al.}(2016)Stange, Wirths, Geiger, Schulte-Braucks,
  Marzban, von~den Driesch, Mussler, Zabel, Stoica, Hartmann, Mantl, Ikonic,
  GrŸtzmacher, Sigg, Witzens, and Buca]{Stange:2016ip}
Stange,~D. \latin{et~al.}  Optically {Pumped} {GeSn} {Microdisk} {Lasers} on
  {Si}. \emph{ACS Photonics} \textbf{2016}, \emph{3}, 1279--1285\relax
\mciteBstWouldAddEndPuncttrue
\mciteSetBstMidEndSepPunct{\mcitedefaultmidpunct}
{\mcitedefaultendpunct}{\mcitedefaultseppunct}\relax
\EndOfBibitem
\bibitem[Reboud \latin{et~al.}(2017)Reboud, Gassenq, Pauc, Aubin, Milord, Thai,
  Bertrand, Guilloy, Rouchon, Rothman, Zabel, Armand~Pilon, Sigg, Chelnokov,
  Hartmann, and Calvo]{Reboud:2017cr}
Reboud,~V. \latin{et~al.}  Optically Pumped {GeSn} Micro-Disks with 16\% {Sn}
  Lasing at 3.1 $\mu$m up to 180 {K}. \emph{Appl. Phys. Lett.} \textbf{2017},
  \emph{111}, 092101\relax
\mciteBstWouldAddEndPuncttrue
\mciteSetBstMidEndSepPunct{\mcitedefaultmidpunct}
{\mcitedefaultendpunct}{\mcitedefaultseppunct}\relax
\EndOfBibitem
\bibitem[Al-Kabi \latin{et~al.}(2016)Al-Kabi, Ghetmiri, Margetis, Pham, Zhou,
  Dou, Collier, Quinde, Du, Mosleh, Liu, Sun, Soref, Tolle, Li, Mortazavi,
  Naseem, and Yu]{AlKabi:2016de}
Al-Kabi,~S. \latin{et~al.}  An Optically Pumped 2.5 $\mu$m {GeSn} Laser on {Si}
  Operating at 110 {K}. \emph{Appl. Phys. Lett.} \textbf{2016}, \emph{109},
  171105\relax
\mciteBstWouldAddEndPuncttrue
\mciteSetBstMidEndSepPunct{\mcitedefaultmidpunct}
{\mcitedefaultendpunct}{\mcitedefaultseppunct}\relax
\EndOfBibitem
\bibitem[Margetis \latin{et~al.}(2018)Margetis, Al-Kabi, Du, Dou, Zhou, Pham,
  Grant, Ghetmiri, Mosleh, Li, Liu, Sun, Soref, Tolle, Mortazavi, and
  Yu]{Margetis:2017fs}
Margetis,~J. \latin{et~al.}  Si-{Based} {GeSn} {Lasers} with {Wavelength}
  {Coverage} of 2Ð3 $\mu$m and {Operating} {Temperatures} up to 180 {K}.
  \emph{ACS Photonics} \textbf{2018}, \emph{5}, 827--833\relax
\mciteBstWouldAddEndPuncttrue
\mciteSetBstMidEndSepPunct{\mcitedefaultmidpunct}
{\mcitedefaultendpunct}{\mcitedefaultseppunct}\relax
\EndOfBibitem
\bibitem[Dou \latin{et~al.}(2018)Dou, Zhou, Margetis, Ghetmiri, Al-Kabi, Du,
  Liu, Sun, Soref, Tolle, Li, Mortazavi, and Yu]{Li:2018ka}
Dou,~W.; Zhou,~Y.; Margetis,~J.; Ghetmiri,~S.~A.; Al-Kabi,~S.; Du,~W.; Liu,~J.;
  Sun,~G.; Soref,~R.~A.; Tolle,~J.; Li,~B.; Mortazavi,~M.; Yu,~S.-Q. Optically
  Pumped Lasing at 3 $\mu$m from Compositionally Graded {GeSn} with Tin up to
  22.3\%. \emph{Opt. Lett.} \textbf{2018}, \emph{43}, 4558--4561\relax
\mciteBstWouldAddEndPuncttrue
\mciteSetBstMidEndSepPunct{\mcitedefaultmidpunct}
{\mcitedefaultendpunct}{\mcitedefaultseppunct}\relax
\EndOfBibitem
\bibitem[Kurosawa \latin{et~al.}(2015)Kurosawa, Kato, Yamaha, Taoka, Nakatsuka,
  and Zaima]{Kurosawa:2015ba}
Kurosawa,~M.; Kato,~M.; Yamaha,~T.; Taoka,~N.; Nakatsuka,~O.; Zaima,~S.
  {Near-infrared light absorption by polycrystalline SiSn alloys grown on
  insulating layers}. \emph{Applied Physics Letters} \textbf{2015}, \emph{106},
  171908\relax
\mciteBstWouldAddEndPuncttrue
\mciteSetBstMidEndSepPunct{\mcitedefaultmidpunct}
{\mcitedefaultendpunct}{\mcitedefaultseppunct}\relax
\EndOfBibitem
\bibitem[Tolle \latin{et~al.}(2006)Tolle, Chizmeshya, Fang, Kouvetakis,
  D{\textquoteright}Costa, Hu, Men{\'e}ndez, and Tsong]{Tolle:2006js}
Tolle,~J.; Chizmeshya,~A. V.~G.; Fang,~Y.~Y.; Kouvetakis,~J.;
  D{\textquoteright}Costa,~V.~R.; Hu,~C.~W.; Men{\'e}ndez,~J.; Tsong,~I. S.~T.
  {Low temperature chemical vapor deposition of Si-based compounds via
  SiH3SiH2SiH3: Metastable SiSn$\slash$GeSn$\slash$Si(100) heteroepitaxial
  structures}. \emph{Applied Physics Letters} \textbf{2006}, \emph{89},
  231924\relax
\mciteBstWouldAddEndPuncttrue
\mciteSetBstMidEndSepPunct{\mcitedefaultmidpunct}
{\mcitedefaultendpunct}{\mcitedefaultseppunct}\relax
\EndOfBibitem
\bibitem[McCarthy \latin{et~al.}(2021)McCarthy, Ju, Schaefer, Yu, and
  Zhang]{10.1063/5.0063179}
McCarthy,~T.~T.; Ju,~Z.; Schaefer,~S.; Yu,~S.-Q.; Zhang,~Y.-H.
  {Momentum(k)-space carrier separation using SiGeSn alloys for photodetector
  applications}. \emph{Journal of Applied Physics} \textbf{2021}, \emph{130},
  223102\relax
\mciteBstWouldAddEndPuncttrue
\mciteSetBstMidEndSepPunct{\mcitedefaultmidpunct}
{\mcitedefaultendpunct}{\mcitedefaultseppunct}\relax
\EndOfBibitem
\bibitem[Moody \latin{et~al.}(2014)Moody, Ceguerra, Breen, Cui, Gault,
  Stephenson, Marceau, Powles, and Ringer]{10.1038/ncomms6501}
Moody,~M.~P.; Ceguerra,~A.~V.; Breen,~A.~J.; Cui,~X.~Y.; Gault,~B.;
  Stephenson,~L.~T.; Marceau,~R. K.~W.; Powles,~R.~C.; Ringer,~S.~P.
  {Atomically resolved tomography to directly inform simulations for
  structure-property relationships}. \emph{Nature Communications}
  \textbf{2014}, \emph{5}, 1 -- 10\relax
\mciteBstWouldAddEndPuncttrue
\mciteSetBstMidEndSepPunct{\mcitedefaultmidpunct}
{\mcitedefaultendpunct}{\mcitedefaultseppunct}\relax
\EndOfBibitem
\bibitem[Zhang \latin{et~al.}(2017)Zhang, Zhao, Jin, Xue, Velisa, Bei, Huang,
  Ko, Pagan, Neuefeind, Weber, and Zhang]{Zhang:2017bw}
Zhang,~F.; Zhao,~S.; Jin,~K.; Xue,~H.; Velisa,~G.; Bei,~H.; Huang,~R.; Ko,~J.;
  Pagan,~D.; Neuefeind,~J.; Weber,~W.; Zhang,~Y. Local {Structure} and
  {Short}-{Range} {Order} in a {NiCoCr} {Solid} {Solution} {Alloy}. \emph{Phys.
  Rev. Lett.} \textbf{2017}, \emph{118}, 205501\relax
\mciteBstWouldAddEndPuncttrue
\mciteSetBstMidEndSepPunct{\mcitedefaultmidpunct}
{\mcitedefaultendpunct}{\mcitedefaultseppunct}\relax
\EndOfBibitem
\bibitem[Mukherjee \latin{et~al.}(2017)Mukherjee, Kodali, Isheim, Wirths,
  Hartmann, Buca, Seidman, and Moutanabbir]{Mukherjee:2017is}
Mukherjee,~S.; Kodali,~N.; Isheim,~D.; Wirths,~S.; Hartmann,~J.~M.; Buca,~D.;
  Seidman,~D.~N.; Moutanabbir,~O. Short-Range Atomic Ordering in Nonequilibrium
  Silicon-Germanium-Tin Semiconductors. \emph{Phys. Rev. B} \textbf{2017},
  \emph{95}, 161402\relax
\mciteBstWouldAddEndPuncttrue
\mciteSetBstMidEndSepPunct{\mcitedefaultmidpunct}
{\mcitedefaultendpunct}{\mcitedefaultseppunct}\relax
\EndOfBibitem
\bibitem[Shimura \latin{et~al.}(2017)Shimura, Asano, Yamaha, Fukuda, Takeuchi,
  Nakatsuka, and Zaima]{Shimura:2017gb}
Shimura,~Y.; Asano,~T.; Yamaha,~T.; Fukuda,~M.; Takeuchi,~W.; Nakatsuka,~O.;
  Zaima,~S. {EXAFS study of local structure contributing to Sn stability in
  SiyGe1-y-zSnz}. \emph{Materials Science in Semiconductor Processing}
  \textbf{2017}, \emph{70}, 133--138\relax
\mciteBstWouldAddEndPuncttrue
\mciteSetBstMidEndSepPunct{\mcitedefaultmidpunct}
{\mcitedefaultendpunct}{\mcitedefaultseppunct}\relax
\EndOfBibitem
\bibitem[Sher \latin{et~al.}(1987)Sher, Schilfgaarde, Chen, and
  Chen]{10.1103/physrevb.36.4279}
Sher,~A.; Schilfgaarde,~M.~v.; Chen,~A.-B.; Chen,~W. {Quasichemical
  approximation in binary alloys}. \emph{Physical Review B} \textbf{1987},
  \emph{36}, 4279 -- 4295\relax
\mciteBstWouldAddEndPuncttrue
\mciteSetBstMidEndSepPunct{\mcitedefaultmidpunct}
{\mcitedefaultendpunct}{\mcitedefaultseppunct}\relax
\EndOfBibitem
\bibitem[Manzoor \latin{et~al.}(2018)Manzoor, Pandey, Chakraborty, Phillpot,
  and Aidhy]{Manzoor:2018er}
Manzoor,~A.; Pandey,~S.; Chakraborty,~D.; Phillpot,~S.~R.; Aidhy,~D.~S.
  {Entropy contributions to phase stability in binary random solid solutions}.
  \emph{npj Computational Materials} \textbf{2018}, \emph{4}, 1 -- 10\relax
\mciteBstWouldAddEndPuncttrue
\mciteSetBstMidEndSepPunct{\mcitedefaultmidpunct}
{\mcitedefaultendpunct}{\mcitedefaultseppunct}\relax
\EndOfBibitem
\bibitem[Esters \latin{et~al.}(2021)Esters, Oses, Hicks, Mehl, Jahn‡tek,
  Hossain, Maria, Brenner, Toher, and Curtarolo]{10.1038/s41467-021-25979-5}
Esters,~M.; Oses,~C.; Hicks,~D.; Mehl,~M.~J.; Jahn‡tek,~M.; Hossain,~M.~D.;
  Maria,~J.-P.; Brenner,~D.~W.; Toher,~C.; Curtarolo,~S. {Settling the matter
  of the role of vibrations in the stability of high-entropy carbides}.
  \emph{Nature Communications} \textbf{2021}, \emph{12}, 5747\relax
\mciteBstWouldAddEndPuncttrue
\mciteSetBstMidEndSepPunct{\mcitedefaultmidpunct}
{\mcitedefaultendpunct}{\mcitedefaultseppunct}\relax
\EndOfBibitem
\bibitem[Moutanabbir \latin{et~al.}(2021)Moutanabbir, Assali, Gong, O'Reilly,
  Broderick, Marzban, Witzens, Du, Yu, Chelnokov, Buca, and
  Nam]{10.1063/5.0043511}
Moutanabbir,~O.; Assali,~S.; Gong,~X.; O'Reilly,~E.; Broderick,~C.~A.;
  Marzban,~B.; Witzens,~J.; Du,~W.; Yu,~S.-Q.; Chelnokov,~A.; Buca,~D.; Nam,~D.
  {Monolithic infrared silicon photonics: The rise of (Si)GeSn semiconductors}.
  \emph{Applied Physics Letters} \textbf{2021}, \emph{118}, 110502\relax
\mciteBstWouldAddEndPuncttrue
\mciteSetBstMidEndSepPunct{\mcitedefaultmidpunct}
{\mcitedefaultendpunct}{\mcitedefaultseppunct}\relax
\EndOfBibitem
\bibitem[Mieghem(1992)]{10.1103/revmodphys.64.755}
Mieghem,~P.~V. {Theory of band tails in heavily doped semiconductors}.
  \emph{Reviews of Modern Physics} \textbf{1992}, \emph{64}, 755--793\relax
\mciteBstWouldAddEndPuncttrue
\mciteSetBstMidEndSepPunct{\mcitedefaultmidpunct}
{\mcitedefaultendpunct}{\mcitedefaultseppunct}\relax
\EndOfBibitem
\bibitem[Moutanabbir and Gšsele(2010)Moutanabbir, and
  Gšsele]{10.1146/annurev-matsci-070909-104448}
Moutanabbir,~O.; Gšsele,~U. {Heterogeneous Integration of Compound
  Semiconductors}. \emph{Annual Review of Materials Research} \textbf{2010},
  \emph{40}, 469--500\relax
\mciteBstWouldAddEndPuncttrue
\mciteSetBstMidEndSepPunct{\mcitedefaultmidpunct}
{\mcitedefaultendpunct}{\mcitedefaultseppunct}\relax
\EndOfBibitem
\bibitem[Ding \latin{et~al.}(2019)Ding, Zhang, Chen, Fu, Chen, Chen, Gu, Wei,
  Bei, Gao, Wen, Li, Zhang, Zhu, Ritchie, and Yu]{10.1038/s41586-019-1617-1}
Ding,~Q. \latin{et~al.}  {Tuning element distribution, structure and properties
  by composition in high-entropy alloys}. \emph{Nature} \textbf{2019},
  \emph{574}, 223 -- 227\relax
\mciteBstWouldAddEndPuncttrue
\mciteSetBstMidEndSepPunct{\mcitedefaultmidpunct}
{\mcitedefaultendpunct}{\mcitedefaultseppunct}\relax
\EndOfBibitem
\bibitem[Metropolis \latin{et~al.}(1953)Metropolis, Rosenbluth, Rosenbluth,
  Teller, and Teller]{Metropolis:1953in}
Metropolis,~N.; Rosenbluth,~A.~W.; Rosenbluth,~M.~N.; Teller,~A.~H.; Teller,~E.
  Equation of {State} {Calculations} by {Fast} {Computing} {Machines}. \emph{J.
  Chem. Phys.} \textbf{1953}, \emph{21}, 1087--1092\relax
\mciteBstWouldAddEndPuncttrue
\mciteSetBstMidEndSepPunct{\mcitedefaultmidpunct}
{\mcitedefaultendpunct}{\mcitedefaultseppunct}\relax
\EndOfBibitem
\bibitem[Kresse and Hafner(1993)Kresse, and Hafner]{Kresse:1993ty}
Kresse,~G.; Hafner,~J. Ab Initio Molecular Dynamics for Liquid Metals.
  \emph{Phys. Rev. B} \textbf{1993}, \emph{47}, 558--561\relax
\mciteBstWouldAddEndPuncttrue
\mciteSetBstMidEndSepPunct{\mcitedefaultmidpunct}
{\mcitedefaultendpunct}{\mcitedefaultseppunct}\relax
\EndOfBibitem
\bibitem[Kresse and Joubert(1999)Kresse, and Joubert]{Kresse:1999tq}
Kresse,~G.; Joubert,~D. From Ultrasoft Pseudopotentials to the Projector
  Augmented-Wave Method. \emph{Phys. Rev. B} \textbf{1999}, \emph{59},
  1758--1775\relax
\mciteBstWouldAddEndPuncttrue
\mciteSetBstMidEndSepPunct{\mcitedefaultmidpunct}
{\mcitedefaultendpunct}{\mcitedefaultseppunct}\relax
\EndOfBibitem
\bibitem[Kresse and FurthmŸller(1996)Kresse, and FurthmŸller]{Kresse:1996kg}
Kresse,~G.; FurthmŸller,~J. Efficiency of Ab-Initio Total Energy Calculations
  for Metals and Semiconductors Using a Plane-Wave Basis Set.
  \emph{Computational Materials Science} \textbf{1996}, \emph{6}, 15--50\relax
\mciteBstWouldAddEndPuncttrue
\mciteSetBstMidEndSepPunct{\mcitedefaultmidpunct}
{\mcitedefaultendpunct}{\mcitedefaultseppunct}\relax
\EndOfBibitem
\bibitem[Kresse and FurthmŸller(1996)Kresse, and FurthmŸller]{Kresse:1996vf}
Kresse,~G.; FurthmŸller,~J. Efficient Iterative Schemes for Ab Initio
  Total-Energy Calculations Using a Plane-Wave Basis Set. \emph{Phys. Rev. B}
  \textbf{1996}, \emph{54}, 11169--11186\relax
\mciteBstWouldAddEndPuncttrue
\mciteSetBstMidEndSepPunct{\mcitedefaultmidpunct}
{\mcitedefaultendpunct}{\mcitedefaultseppunct}\relax
\EndOfBibitem
\bibitem[Ceperley and Alder(1980)Ceperley, and Alder]{Ceperley:1980gc}
Ceperley,~D.~M.; Alder,~B.~J. Ground {State} of the {Electron} {Gas} by a
  {Stochastic} {Method}. \emph{Phys. Rev. Lett.} \textbf{1980}, \emph{45},
  566--569\relax
\mciteBstWouldAddEndPuncttrue
\mciteSetBstMidEndSepPunct{\mcitedefaultmidpunct}
{\mcitedefaultendpunct}{\mcitedefaultseppunct}\relax
\EndOfBibitem
\bibitem[Eckhardt \latin{et~al.}(2014)Eckhardt, Hummer, and
  Kresse]{Eckhardt:2014gz}
Eckhardt,~C.; Hummer,~K.; Kresse,~G. Indirect-To-Direct Gap Transition in
  Strained and Unstrained Sn$_x$Ge$_{1-x}$ Alloys. \emph{Phys. Rev. B}
  \textbf{2014}, \emph{89}, 165201\relax
\mciteBstWouldAddEndPuncttrue
\mciteSetBstMidEndSepPunct{\mcitedefaultmidpunct}
{\mcitedefaultendpunct}{\mcitedefaultseppunct}\relax
\EndOfBibitem
\bibitem[Polak \latin{et~al.}(2017)Polak, Scharoch, and
  Kudrawiec]{Polak:2017fh}
Polak,~M.~P.; Scharoch,~P.; Kudrawiec,~R. The Electronic Band Structure of
  Ge$_{\text{1-x}}$Sn$_{\text{x}}$ in the Full Composition Range: Indirect,
  Direct, and Inverted Gaps Regimes, Band Offsets, and the {Burstein}Ð{Moss}
  Effect. \emph{J. Phys. D: Appl. Phys.} \textbf{2017}, \emph{50}, 195103\relax
\mciteBstWouldAddEndPuncttrue
\mciteSetBstMidEndSepPunct{\mcitedefaultmidpunct}
{\mcitedefaultendpunct}{\mcitedefaultseppunct}\relax
\EndOfBibitem
\bibitem[Haas \latin{et~al.}(2009)Haas, Tran, and Blaha]{Haas:2009be}
Haas,~P.; Tran,~F.; Blaha,~P. Calculation of the Lattice Constant of Solids
  with Semilocal Functionals. \emph{Phys. Rev. B} \textbf{2009}, \emph{79},
  085104\relax
\mciteBstWouldAddEndPuncttrue
\mciteSetBstMidEndSepPunct{\mcitedefaultmidpunct}
{\mcitedefaultendpunct}{\mcitedefaultseppunct}\relax
\EndOfBibitem
\bibitem[Tran and Blaha(2009)Tran, and Blaha]{Tran:2009kk}
Tran,~F.; Blaha,~P. Accurate {Band} {Gaps} of {Semiconductors} and {Insulators}
  with a {Semilocal} {Exchange}-{Correlation} {Potential}. \emph{Phys. Rev.
  Lett.} \textbf{2009}, \emph{102}, 226401\relax
\mciteBstWouldAddEndPuncttrue
\mciteSetBstMidEndSepPunct{\mcitedefaultmidpunct}
{\mcitedefaultendpunct}{\mcitedefaultseppunct}\relax
\EndOfBibitem
\bibitem[Monkhorst and Pack(1976)Monkhorst, and Pack]{Monkhorst:1976cv}
Monkhorst,~H.~J.; Pack,~J.~D. {Special points for Brillouin-zone integrations}.
  \emph{Physical Review B} \textbf{1976}, \emph{13}, 5188\relax
\mciteBstWouldAddEndPuncttrue
\mciteSetBstMidEndSepPunct{\mcitedefaultmidpunct}
{\mcitedefaultendpunct}{\mcitedefaultseppunct}\relax
\EndOfBibitem
\bibitem[Popescu and Zunger(2010)Popescu, and Zunger]{Popescu:2010jd}
Popescu,~V.; Zunger,~A. Effective {Band} {Structure} of {Random} {Alloys}.
  \emph{Phys. Rev. Lett.} \textbf{2010}, \emph{104}, 236403\relax
\mciteBstWouldAddEndPuncttrue
\mciteSetBstMidEndSepPunct{\mcitedefaultmidpunct}
{\mcitedefaultendpunct}{\mcitedefaultseppunct}\relax
\EndOfBibitem
\bibitem[Rubel \latin{et~al.}(2014)Rubel, Bokhanchuk, Ahmed, and
  Assmann]{Rubel:2014fv}
Rubel,~O.; Bokhanchuk,~A.; Ahmed,~S.~J.; Assmann,~E. Unfolding the Band
  Structure of Disordered Solids: {From} Bound States to High-Mobility {Kane}
  Fermions. \emph{Phys. Rev. B} \textbf{2014}, \emph{90}, 115202\relax
\mciteBstWouldAddEndPuncttrue
\mciteSetBstMidEndSepPunct{\mcitedefaultmidpunct}
{\mcitedefaultendpunct}{\mcitedefaultseppunct}\relax
\EndOfBibitem
\bibitem[Dove(1993)]{Dove_1993}
Dove,~M.~T. \emph{Introduction to Lattice Dynamics}; Cambridge Topics in
  Mineral Physics and Chemistry; Cambridge University Press, 1993\relax
\mciteBstWouldAddEndPuncttrue
\mciteSetBstMidEndSepPunct{\mcitedefaultmidpunct}
{\mcitedefaultendpunct}{\mcitedefaultseppunct}\relax
\EndOfBibitem
\bibitem[Togo and Tanaka(2015)Togo, and
  Tanaka]{10.1016/j.scriptamat.2015.07.021}
Togo,~A.; Tanaka,~I. {First principles phonon calculations in materials
  science}. \emph{Scripta Materialia} \textbf{2015}, \emph{108}, 1--5\relax
\mciteBstWouldAddEndPuncttrue
\mciteSetBstMidEndSepPunct{\mcitedefaultmidpunct}
{\mcitedefaultendpunct}{\mcitedefaultseppunct}\relax
\EndOfBibitem
\end{mcitethebibliography}
\providecommand{\latin}[1]{#1}
\makeatletter
\providecommand{\doi}
  {\begingroup\let\do\@makeother\dospecials
  \catcode`\{=1 \catcode`\}=2 \doi@aux}
\providecommand{\doi@aux}[1]{\endgroup\texttt{#1}}
\makeatother
\providecommand*\mcitethebibliography{\thebibliography}
\csname @ifundefined\endcsname{endmcitethebibliography}
  {\let\endmcitethebibliography\endthebibliography}{}

\end{document}